\documentclass{att-2014}
\usepackage{epsfig,url,graphicx,microtype}

\begin{document}

\title{JADE, TraSMAPI and SUMO: A tool-chain for simulating traffic light control}

\author{
Tiago M. L. Azevedo
\quad 
Paulo J. M. de Ara\'{u}jo
\quad
Rosaldo J. F. Rossetti
\quad
Ana Paula C. Rocha
\\
Artificial Intelligence and Computer Science Lab\\
Department of Informatics Engineering\\
Faculty of Engineering, University of Porto, Portugal\\
\email{\{tiago.manuel, paulo.araujo, rossetti, arocha\}@fe.up.pt}
}

\maketitle

\begin{abstract}
Increased stress, fuel consumption, air pollution, accidents and delays are some of the consequences of traffic congestion usually incurring in tremendous economic impacts, which society aims to remedy in order to leverage a sustainable development. Recently, unconventional means for modeling and controlling such complex traffic systems relying on multi-agent systems have arisen. This paper contributes to the understanding of such complex and highly dynamic systems by proposing an open-source tool-chain to implement multi-agent-based solutions in traffic and transportation. The proposed approach relies on two very popular tools in both domains, with focus on traffic light control. This tool-chain consists in combining JADE (\textbf{J}ava \textbf{A}gent \textbf{DE}velopment Framework), for the implementation of multi-agent systems, with SUMO (\textbf{S}imulation of \textbf{U}rban \textbf{MO}bility), for the microscopic simulation of traffic interactions. TraSMAPI (\textbf{Tra}ffic \textbf{S}imulation \textbf{M}anager \textbf{A}pplication \textbf{P}rogramming \textbf{I}nterface) is used to combine JADE and SUMO allowing communication between them. A demonstration of the concept is presented to illustrate the main features of this tool-chain, using Q-Learning as the reinforcement learning method for each traffic light agent in a simulated network. Results demonstrate the feasibility of the proposed framework as a practical means to experiment with different agent-based designs of intelligent transportation solutions.
\end{abstract}







\section{Introduction}
Nowadays urban centers face the daily problem of traffic congestion, which in addition to the obvious confusion can create also other negative consequences. Increased stress, fuel consumption, air pollution, accidents and delays are some of these consequences, which society aims to remedy in order to leverage a sustainable development, while mitigating tremendous economic impacts.

Solutions to this problem have evolved over time, more in an immediate response perspective than on a long-term resolution perspective. Initially, the approach was based on the construction of alternative routes with increased capacity. However, available money and territorial area ceased to exist for continuing implementation of this sort of solution. In parallel, traffic lights and roundabouts were introduced but the urban centers continued growth now are demanding more advanced and efficient alternative measures.

The aim of the work described in this paper was to use a tool-chain that allows us to implement a multi-agent system (MAS) for traffic light control. Therefore, a multi-agent system approach was used to answer the daily problem of traffic congestion. This tool-chain consisted in integrating JADE (\textbf{J}ava \textbf{A}gent \textbf{DE}velopment Framework) for controlling the multi-agent system to SUMO (\textbf{S}imulation of \textbf{U}rban \textbf{MO}bility) for traffic simulation. TraSMAPI (\textbf{Tra}ffic \textbf{S}imulation \textbf{M}anager \textbf{A}pplication \textbf{P}rogramming \textbf{I}nterface) was the middleware combining JADE and SUMO and allowing communication between both environments. For the sake of illustration, the implemented agents' learning method was Q-Learning.

As a motivation, just a few simulation tools truly support the concept of agents and multi-agent systems in traffic simulation; MATSim-T \cite{balmer2008agent, balmer2009matsim} and ITSUMO \cite{da2006itsumo, bazzan2010itsumo} are good examples to be mentioned. However, no standard of wide reach for the implementation of such tools actually exists. Indeed, alternative approaches would require either general purpose MAS-based simulators to be adapted to the specific domain of traffic and transportation, or the other way around with the adaptation of traffic simulators to be adapted so as to support the MAS-based models. With our approach, we expect to benefit from both worlds on an integrated basis. Also, it is important to notice that although SUMO and ITSUMO are both open-source microscopic simulators and have a quite similar acronym, they are no related applications. ITSUMO is a Cellular-Automata-based simulator, whereas SUMO uses a continuous representation of space on road segments. Besides, ITSUMO explicitly consider the metaphor of agents, whereas SUMO can be considered a traditional microscopic simulator, where agents are not explicitly implemented.

The expected contribution of this work, rather than implementing a new agent-based simulator from scratch, adapting or extending existing ones, is to devise an open-source tool-chain to implement MAS-T (MAS in traffic and transportation) on the basis of two very popular tools in both domains. On the one hand, JADE supports the implementation of any MAS solution and, on the other hand, SUMO supports an appropriate representation of the traffic environment in which agents inhabit and perform their tasks.

This paper will start to deeply describe the tools. The conceived model is detailed and instantiated in the proposed tool-chain. An experimental set-up is used to illustrate the proposed approach, followed by the discussion of preliminary results. After discussion on related works, conclusions are drawn as well as are further developments suggested.

\section{A MAS-based traffic simulation tool-chain}
The MAS-based traffic simulation tool-chain used consisted in three main tools: JADE, SUMO and TraSMAPI.

A multi-agent system based approach seems to be the appropriate way to represent the different traffic lights in a network. Consequently, it is necessary that a multi-agent system framework take care of the different agent behaviours, as it is the case in JADE.

Next, a microscopic simulator is needed to take care of the traffic road dynamics, such as vehicles decisions. It should be noted that although it is necessary to have vehicles in order to test traffic light control, these vehicles do not need to be modeled as agents. It would be very computationally expensive to simulate a huge quantity of vehicles, each one with driver's decision-making and other cognitive aspects and details. SUMO was the microscopic simulator chosen.

Finally, as traffic lights are considered to be agents, it is necessary they communicate with the simulator. This is important so as to allow their traffic lights in the simulation to have the semaphore plans always updated and agents to perceive the network dynamics. This communication was made through TraSMAPI, consisting of an integration API implemented in Java.

\subsection{JADE}
JADE is a framework completely developed in Java. It simplifies the implementation of multi-agent systems through a middleware that complies with the FIPA\footnote{\textbf{F}oundation of \textbf{I}ntelligent \textbf{P}hysical \textbf{A}gents, an organization that promotes agent-based technology and the interoperability of its standards with other technologies} specifications and through a set of graphical tools that supports the debugging and deployment phases. The agent platform can be distributed across machines and the configuration can be controlled via a remote GUI \cite{jade:website}. The version used in this work was \textit{4.3.0}, released on March 2013.



One advantage of using JADE to implement MAS is its ability to allow run-time visualisation and control of the interactions among agents in the application. As relevant features for this work, some can be pointed that are not directly connected to agents, that is, are independent of the applications: message transportation, codification and parsing of messages or lifetime of an agent, for instance.

\subsection{SUMO}
SUMO is an open-source program (licenced under GPL\footnote{GNU \textbf{G}eneral \textbf{P}ublic \textbf{L}icense, a free, copyleft license for software and other kinds of works}) for traffic simulation. Its simulation model is microscopic, that is, each vehicle is explicitly modeled, has its own route and moves individually over the network. It is mainly developed by Institute of Transportation Systems, located at German Aerospace Center \cite{sumo:glance}. The version used in this work was \textit{0.18.0}, released on August 2013.

Among other features, it allows the existence of different types of vehicles, roads with several lanes, traffic lights, graphical interface to view the network and the entities that are being simulated, and interoperability with other applications at run-time through an API called TraCI. Moreover, the tool is considered to be fast, still allowing a version without a graphical interface where the simulation is accelerated putting aside visual concerns and overheads\cite{sumo:glance}.

\begin{figure}[h!]
\centering
\includegraphics[width=83mm]{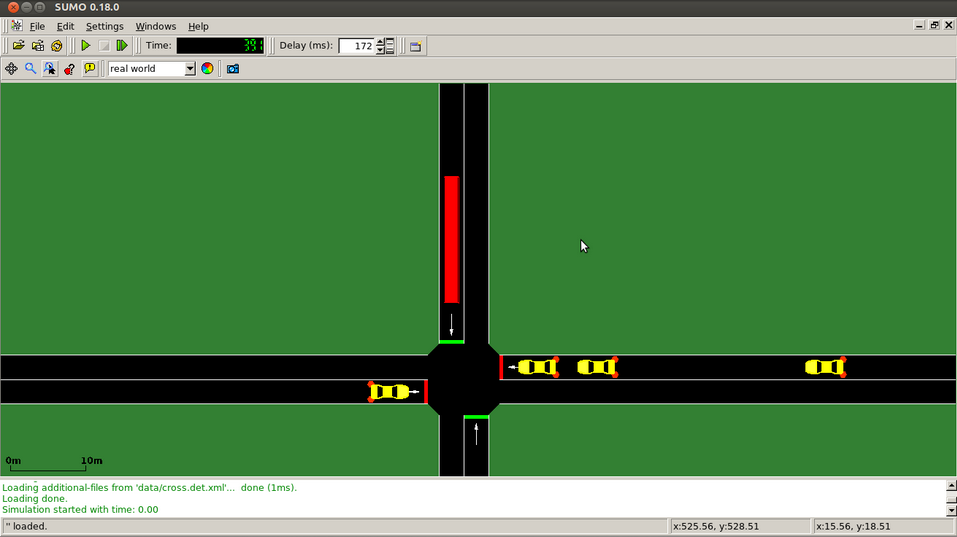}
\caption{SUMO working}
\label{img:sumo_example}
\vspace{-0.9em}
\end{figure}

In Figure \ref{img:sumo_example} it is possible to visualize the SUMO's graphical interface with a running simulation. It is possible to point out almost all specified features: vehicles stopped at the traffic light as well as a long vehicle entering an intersection.

This tool was crucial in this work! First, it allows loading different maps (described in XML files) in order to test various scenarios with vehicles and traffic lights. Then, with the simulation itself there is no need to waste time implementing the dynamics of many vehicles and traffic lights, starting soon with the evaluation of algorithms. Finally, interoperability with other applications allows that each agent can be bound to an entity in SUMO, so that changes in the dynamics of traffic lights, for instance, can be visually seen in the SUMO's graphic interface.

\subsection{TraSMAPI}
TraSMAPI can be seen as a generic API for microscopic traffic that allows real-time communication between agents of urban traffic management (such as vehicles and traffic signals) and the environment created by various simulators. This tool was developed in LIACC (Artificial Intelligence and Computer Science Laboratory), University of Porto, having already been tested with two different simulators, including SUMO \cite{Timoteo2012157}.

\begin{figure}[h!]
\centering
\includegraphics[width=83mm]{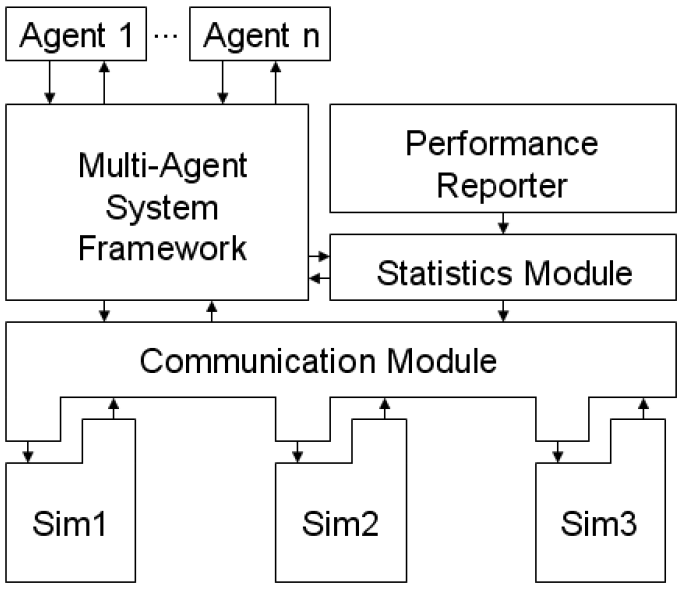}
\caption{TraSMAPI's initial architecture}
\label{img:trasmapi_arch}
\vspace{-0.9em}
\end{figure}

This API offers a higher abstraction level than most of microscopic traffic simulators in such a way that the solution is independent from the microscopic simulator to use. Initially, this tool also aimed to gather relevant metrics/statistics and offer an integrated framework for developing multi-agent systems, as shown in Figure \ref{img:trasmapi_arch} \cite{trasmapi:multisimul}.

As it can be seen, there were three main modules: a communication module with possibility of various microscopic simulators, the module generating statistics and the module for the MAS management. Presently, only the communication module is functional and this is the module that interests to the scope of the presented work.

\subsection{The tool-chain}
In order to achieve a tool-chain with the previous described tools, it was necessary to extend the TraSMAPI API, enabling to build an abstraction over a SUMO's traffic light entity. Thus, it was necessary to implement the communication protocol regarding the methods of traffic light for value retrieval and state change, in TraCI \cite{wegener2008traci}.

\begin{figure}[h!]
\centering
\includegraphics[width=83mm]{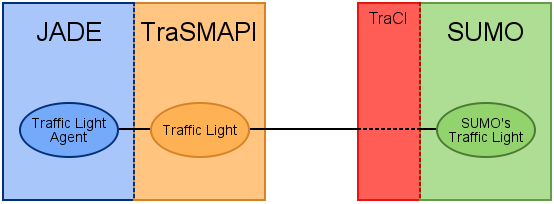}
\caption{Communication between JADE and SUMO using TraSMAPI for a traffic light}
\label{img:simulation_arch}
\vspace{-0.9em}
\end{figure}

The architecture described in Figure \ref{img:simulation_arch} shows how it is possible the existence of one or more traffic light agents. Each traffic light agent has a tie to the respective traffic light to be modeled in SUMO. This tie is supported by the TraSMAPI communication module that interacts with the SUMO API, TraCI.


\section{Experimental Setup}
A simple scenario for the sake of illustration is now described. Although the following scenario is simple and not intended to deeply discuss the appropriateness of implementing traffic control through agents, it illustrates well how our integrated framework could be practically used in this sort of experiments. \\

\subsection{Concepts}
For the purpose of this work, a traffic light is defined as an intersection that has a semaphore plan, which is characterized by a sequence of phases. Each phase has a duration and a color scheme (green, yellow, flashing yellow and/or red), whose values correspond to every possible maneuver at the intersection. The execution of the phases sequence is called a cycle and has a period equal to the sum of the durations of the phases.

\begin{figure}[h!]
\centering
\includegraphics[width=83mm]{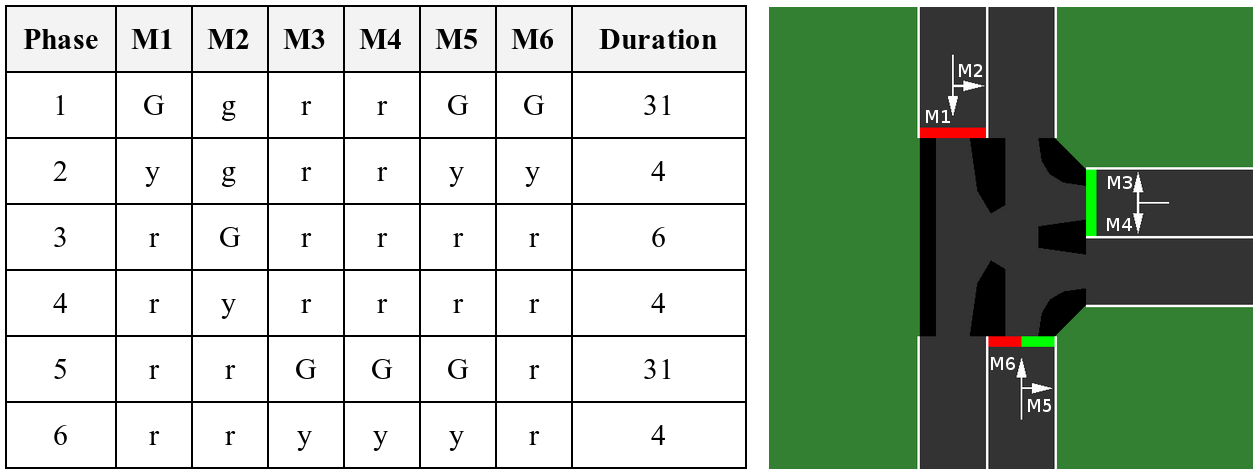}
\caption{Example of a semaphore plan with illustrative image for phase 5}
\label{img:tl_plan_example}
\vspace{-0.9em}
\end{figure}

In Figure \ref{img:tl_plan_example} the intersection has six possible maneuvers, indicated by the arrows, which means that each phase has to specify a color for each maneuver (M1, ..., M6). The sequence of phases is guided by the phase number, and after the end of the sixth phase a 80 cycle duration is completed, following again phase 1. For each maneuver the traffic light may show the green color with symbol \textbf{G}, yellow with symbol \textbf{y}, flashing yellow with symbol \textbf{g} and red with symbol \textbf{r}.

\subsection{Scenario Definition}
\begin{figure}[h!]
\centering
\includegraphics[width=83mm]{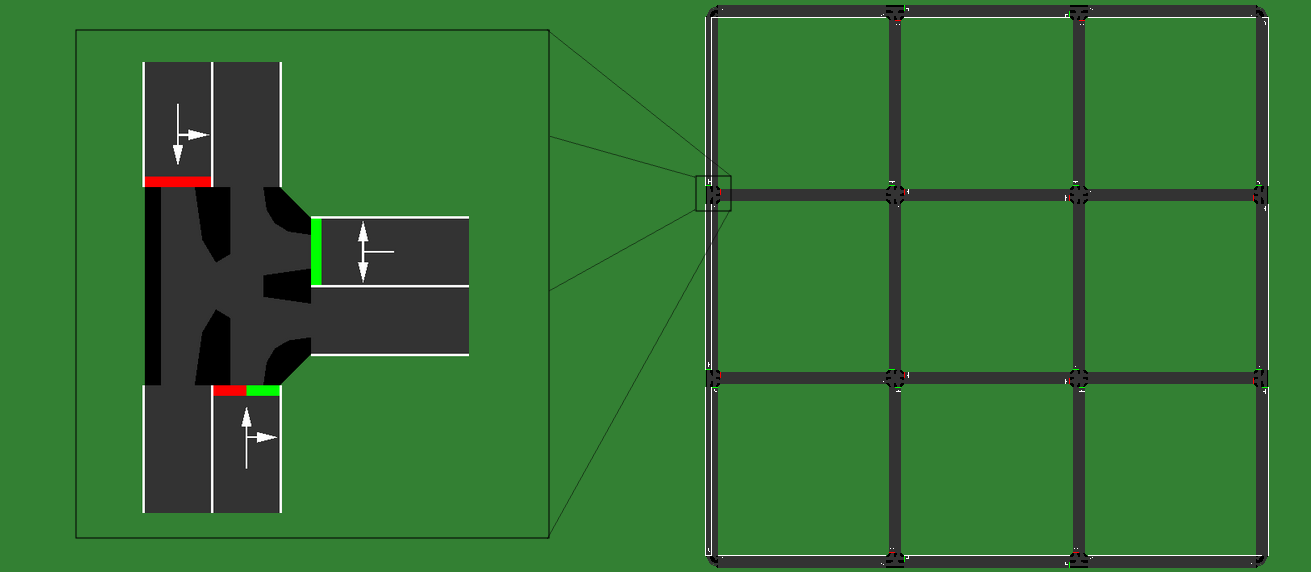}
\caption{The grid map where simulation took place}
\label{img:grid_map}
\vspace{-0.9em}
\end{figure}

As a demonstration of the concept, it was used a grid (Manhattan-like) map (Figure \ref{img:grid_map}) in order to make some experiments for traffic light control. A grid map is relatively simple to implement and where it is fairly possible to define consistent semaphore plans. The Q-learning algorithm was chosen as the learning method for the traffic light agents. 

Thus, these experiments consisted in performing four simulations: one with traffic lights with fixed semaphore plans, one with traffic lights with fixed semaphore plans but with different durations for distinct day periods, another with traffic lights with Q-Learning taking into account the duration of the phases and another with traffic lights with Q-Learning taking into account the duration of the phase and the period of the day. The metrics that will be used to evaluate the results are described in Section \ref{cha:approach_qlearning}. Therefore, to be a basis for comparison, each of the simulations had the same background: the same vehicles leaving at the same time, from the same place and with the same route.

As SUMO's time unit is step (step of execution), and as each step can last more or less a second, it was necessary to make a correlation between number of steps and the time in simulation. This correlation is necessary to implement time compression and allow for an entire day to be simulated correctly and much quicker than in the real-life duration. Thus, the approach taken was that 20000 steps correspond to 1 theoretical simulation hour. There are three traffic scenarios throughout the day: low traffic, medium traffic with a predominance of horizontal flows of vehicles, and heavy traffic. The distribution of traffic is performed according to Table \ref{tab:step_dist}.

\begin{table}
\centering
    \begin{tabular}{|l|l|l|}
    \hline
    Theoretical hour & Starting step & Traffic \\ \hline
    00h00            & 0             & Low     \\
    07h30            & 150 000       & High    \\
    09h00            & 180 000       & Medium  \\
    18h00            & 360 000       & High    \\
    20h00            & 400 000       & Low     \\ \hline
    \end{tabular}
    \caption{Traffic distribution during the day}
    \label{tab:step_dist}
\end{table}

A manual approach was carried out for the definition of the green splitting for the phases in the simulations where Q-Learning was not used. In the specific case of the traffic lights with fixed semaphore plans but with different durations for distinct day periods, in the low traffic period faster green durations were used in opposition to the high traffic period where long green durations were used.
 
Each simulation corresponded to a 4-day simulation. This way, at the end of each simulation, that is, when all vehicles arrived at their destination, metrics were generated.

The tool-chain takes some time to add all desired vehicles at startup. This way, simulation time should not be such that would make the startup take longer than necessary. However, simulation time should be enough so traffic lights have time to learn. 4-day simulation seemed to be the best way for balancing these issues.

It is also important to note that the insertion of network traffic was not made in a distributed manner again because of the slowness that would result with the startup of the tool-chain. Thus, two approaches have been considered for the four simulations: on the one hand, insertions with intervals of 7000 steps, and on the other hand, insertions with intervals of 10000 steps. In each of these intervals, the quantity of vehicles to add would depend on the period of day that the simulation was on. So, in reality, there were 8 simulations.

\subsection{A Q-Learning traffic control}
\label{cha:approach_qlearning}

It is important to be aware that the state representation has influence in the Q-Learning performance, in other words, it is only possible to learn something if it is relevant to the problem. In this sense, it is intended to use two relevant variables: phase durations and period of the day. It is considered that phases initially with duration under 20 seconds will not suffer any variation and the other phases will have durations between 20 and 60 seconds, with a granularity of 5 seconds. Assuming that could exist two or three phases with variable durations for each semaphore plan, there are a total of 81 or 729 duration combinations, respectively. Possible actions are decrease, maintain or increase (-5, 0 or +5 seconds) each variable duration, which results in a Q-Table with 243 or 2187 pairs. Considering the period of the day these numbers would increase.

The reward function consists of two portions: the own reward multiplied by $0,5$ and the weighted average (concerning distance of roads) of neighboring traffic lights rewards multiplied by $0,5$. These rewards are calculated using the average of vehicles in the vicinity of an intersection, during a complete cycle. In what concerns exploration, it is used a 0-greedy strategy. The learning rate was 50\% as well as was the discount factor.

In order to evaluate the learning process, the following metrics will be used:
\begin{itemize}
\itemsep0em
  \item Travel time and average waiting time in queues, that allow to check the individual performance of each vehicle;
  \item Standard deviations of travel times and of average waiting times in queues, that allow to check the network traffic homogeneity, in other words to check whether vehicles will have a similar experience both in travel time and waiting time in queues;
  \item Average of travel times and of average waiting times in queues, that allows to check the global network traffic performance.
\end{itemize}

\subsection{A multi-agent system for traffic control}

System could be implemented using two agent models: an agent for each traffic light with a super coordinator agent, or an agent for each traffic light with distributed coordination. First model allows a greater process synchronization between agents, has a single point of failure for the entire system and has a computation volume highly concentrated in the coordinator. The second model can hardly obtain synchronization but yet in the event of a failure, this is not spread to the entire system, and computation is homogeneous.

\begin{figure*}
\centering
\includegraphics[width=160mm]{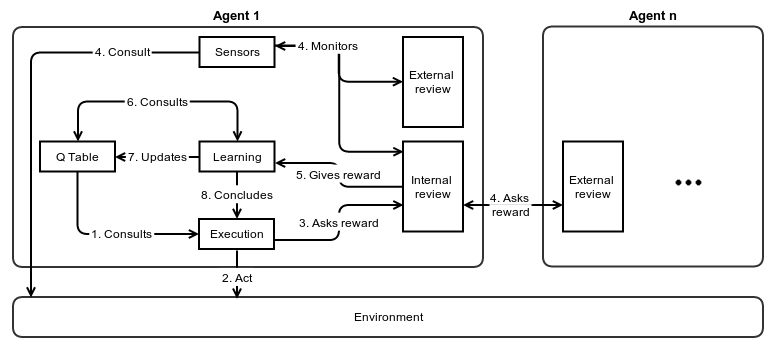}
\caption{Traffic light agent architecture and behaviour}
\label{img:agent_arch}
\vspace{-0.9em}
\end{figure*}

Therefore, system will be implemented using the second model in which agents are traffic lights. The architecture of each agent displayed in Figure \ref{img:agent_arch} is based on a learning agent architecture \cite[p. 54-57]{russell2010artificial} but specified to the Q-Learning process. In this Figure, the presented behaviour does not include the initial phase in which the Q-Table is initialized and where each agent finds the neighbors (in Figure \ref{img:agent_arch} represented as Agent n).

There exist two types of communication between agents: reward requests and answers to reward requests. The former is implemented using the performative \textit{QUERY\_REF} with content "reward", whereas the latter uses the performative \textit{INFORM\_REF} with the reward itself in the content.

\begin{figure}[h!]
\centering
\includegraphics[width=83mm]{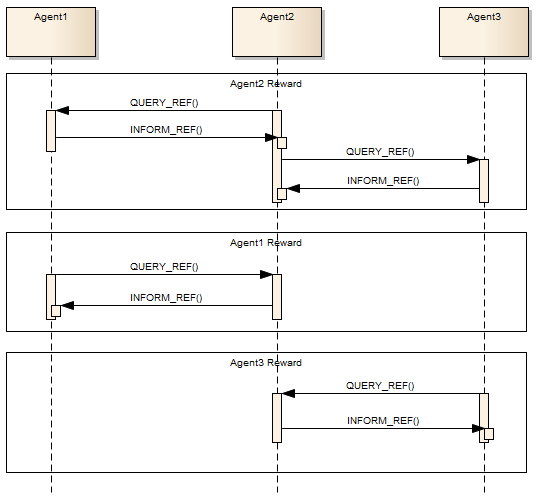}
\caption{Interaction example between agents}
\label{img:agent_interaction}
\vspace{-0.9em}
\end{figure}

Figure \ref{img:agent_interaction} is described a possible situation between agents, in which \textit{Agent2} is a neighbor of \textit{Agent1} and \textit{Agent3}, and \textit{Agent1} and \textit{Agent3} are not neighbors.

\section{Preliminary results and discussion}

\begin{figure*}
\centering
\includegraphics[width=160mm]{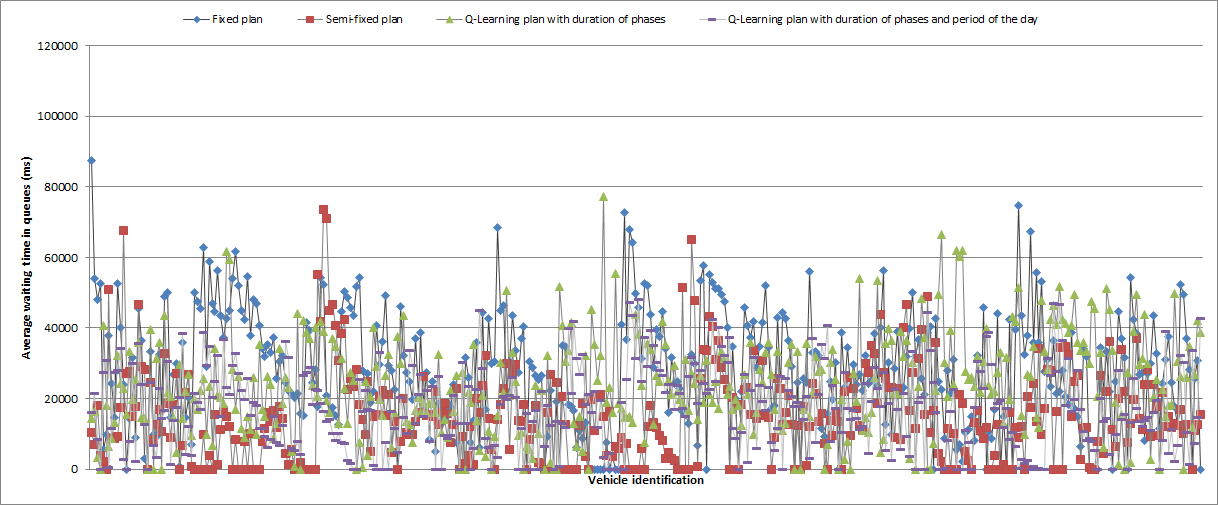}
\caption{The average waiting time in queues with intervals of 10000 steps}
\label{img:waiting_time_10000_single}
\vspace{-0.9em}
\end{figure*}

\begin{figure*}
\centering
\includegraphics[width=160mm]{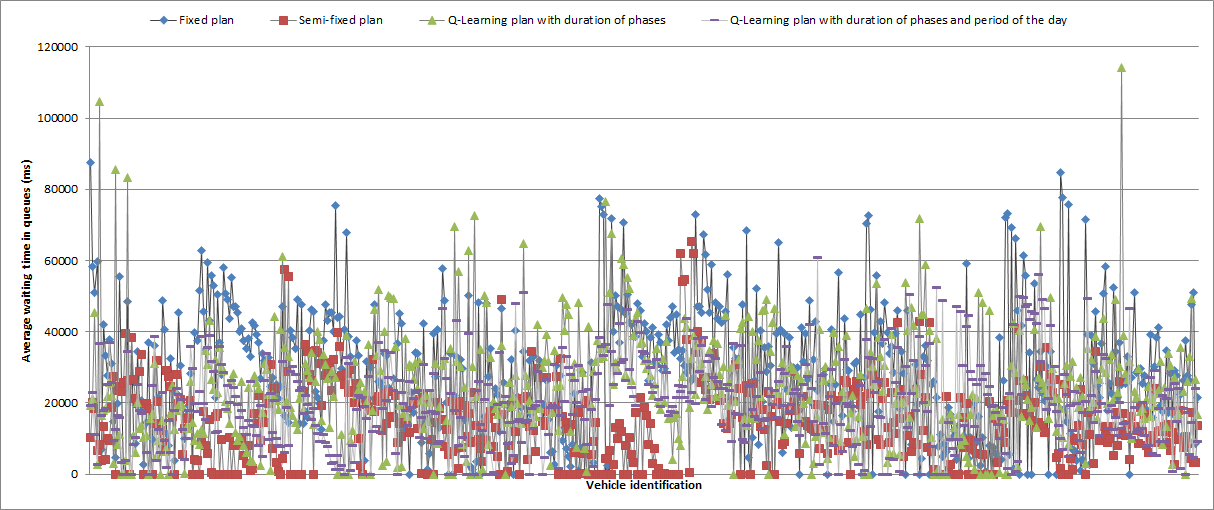}
\caption{The average waiting time in queues with intervals of 7000 steps}
\label{img:waiting_time_7000_single}
\vspace{-0.9em}
\end{figure*}

Figures \ref{img:waiting_time_10000_single} and \ref{img:waiting_time_7000_single} show, for each vehicle, the average waiting time in queues.

For each vehicle, with intervals of 7000 steps there are greater peaks in the average waiting time in queues compared to intervals of 10000 steps. This is explained as the network gets easily saturated with fewer steps and vehicles wait longer in queues. Another fact is that, in individual terms, the average waiting time in queues does not vary a considerably with the types of semaphore plan.




In what concerns the travel time for each vehicle, once again intervals of 7000 steps produce greater peaks, in other words, easily a vehicle takes longer to travel the same path. It is curious to verify that with a changing of the step intervals a vehicle can take longer or shorter in different plans. In other words, unlike the previous metric, there is not a better semaphore plan for the majority of the vehicles, and so a semaphore plan can give better individual results for some vehicles, but not for all vehicles.

\begin{figure}[h!]
\centering
\includegraphics[width=83mm]{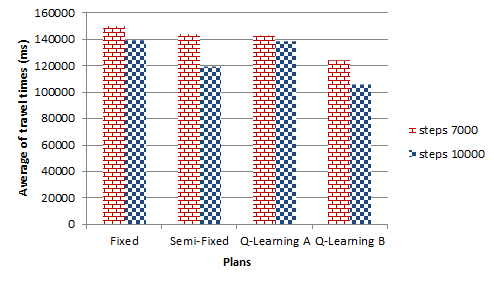}
\caption{Average of travel times}
\label{img:travel_duration_avg}
\vspace{-0.9em}
\end{figure}

\begin{figure}[h!]
\centering
\includegraphics[width=83mm]{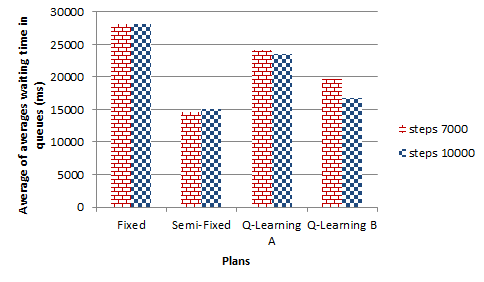}
\caption{Average of averages waiting time in queues}
\label{img:waiting_time_avg}
\vspace{-0.9em}
\end{figure}

Figures \ref{img:travel_duration_avg} and \ref{img:waiting_time_avg} present metrics to a more global evaluation of the explored solutions. It is called \textit{Q-Learning A} to the plan taking into account the duration of the phases and \textit{Q-Learning B} to the plan taking into account the duration of the phases and period of the day.

In these Figures a clear difference between the fixed and semi-fixed plan is shown: while the fixed plan presents the worst results, semi-fixed plans presents the best results, even compared to the Q-Learning plans. Even so, the \textit{Q-Learning B} plan has better results than \textit{Q-Learning A}, as it was expected.

However, what matters the most for the driver is the total travel time. Looking at the Figures, the differences between plans are not big, mainly for the plans with intervals of 7000 steps. Even so, \textit{Q-Learning B} plan has slightly better results.

The peculiar result that semi-fixed plans induces lower waiting times in queues but longer travel times than \textit{Q-Learning B} may be explained. A simple example where this makes sense is that while in \textit{Q-Learning B} a vehicle can pass through a lot of green traffic lights (inducing lower travel times), in the few traffic lights that it has to wait, it waits a lot of time (inducing a greater average waiting time in queues). In the semi-fixed plan a vehicle may have to wait, in average, shorter in queues but as it stops in more traffic lights than in \textit{Q-Learning B}, it takes longer to travel through.

\begin{figure}[h!]
\centering
\includegraphics[width=83mm]{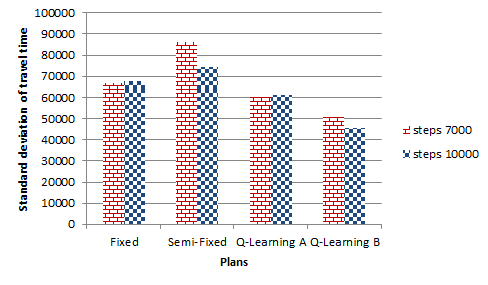}
\caption{Standard deviation of travel time}
\label{img:travel_duration_stdd}
\vspace{-0.9em}
\end{figure}

\begin{figure}[h!]
\centering
\includegraphics[width=83mm]{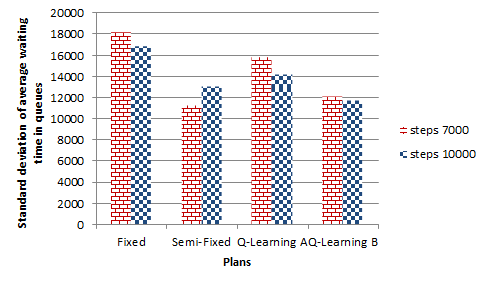}
\caption{Standard deviation of average waiting time in queues}
\label{img:waiting_time_stdd}
\vspace{-0.9em}
\end{figure}

Finally, Figures \ref{img:travel_duration_stdd} and \ref{img:waiting_time_stdd} show the results of standard deviations.

For the averages of waiting times in queues, the semi-fixed plan has the best network traffic homogeneity for intervals of 7000 steps and the second best for intervals of 10000 steps. Nevertheless, in general terms \textit{Q-Learning B} can obtain more network traffic homogeneity.

Passing to the total travel times, \textit{Q-Learning B} can widely overcome the other plans, obtaining greater network traffic homogeneity, both for intervals of 7000 and 10000 steps. The network traffic homogeneity is an important factor for a driver, who intends to know that when he goes to his destination there is not a probability to take longer than it was expected.

\section{Related Works}
The specific case of traffic lights is one of the areas where much has been researched for new solutions, from the design of intersections \cite{koonce2008traffic} (including physical layout and semaphore plans), to the definition of semaphore plans through statistical analysis. Current solutions try to answer the highly dynamic system \cite[p. 343]{chin2012q} using coordinated control. Several methodologies have been used such as genetic algorithms \cite{teklu2007genetic}, fuzzy logic \cite{azimirad2010novel} and reinforcement learning \cite{arel2010reinforcement}.

To date, there are not many solutions for traffic that make full use of the intelligent agent concept. However, the multi-agent system approach has become recognized as a convenient approach for modelling and simulating complex systems \cite{moya2007towards}. Also, it has grown enormously not only applied to traffic but also to transportation in general terms \cite{bazzan2013review}.

In the last decade some microscopic simulators have been developed, such as MITSIM, Paramics, Aimsun, CORSIM and VisSim. However, none of these is strictly defined as agent-based simulation systems, even though they model vehicles in an object-oriented manner. Just a few simulation tools truly support the concept of agents and multi-agent systems in traffic simulation; MATSim-T \cite{balmer2008agent, balmer2009matsim} and ITSUMO \cite{da2006itsumo, bazzan2010itsumo} are good examples to be mentioned.

Regarding this simulation tools some examples of multi-agent system approaches for traffic lights control can be seen in \cite{de2010multi}, \cite{bazzan2010learning}, \cite{lu2011multi} and \cite{de2006reinforcement}. Simulators used in these works were Aimsun, ITSUMO, VisSim and ITSUMO, respectively.

With MAS being recognized as a convenient approach, there must be a sufficiently general way to couple this approach to such a huge quantity of microscopic simulators that exist now. The platform that integrates SUMO and JADE consists of an API intended to allow interoperability among simulators. The platform, coined TraSMAPI, is sufficiently general to allow other simulators to interact with MAS frameworks such JADE. A previous paper \cite{Timoteo2012157} reports on an experiment integrating ITSUMO and SUMO under TraSMAPI, thus demonstrating such an ability. In another study \cite{Vilarinho2014262}, external traffic controller agents operate over Aimsun-simulated scenarios through TraSMAPI. In this specific work, we illustrate how non-agent-based simulators can be extended with TraSMAPI to support MAS-T assessment. There are certainly other options to simulate agent-based traffic and transportation, such as MATSim. Although such tools are open-source then allowing full customisation, the use of JADE over a traditional microscopic simulation tool is expected to promote greater flexibility in terms of agent architectures that can be implemented.

In respect to the described tool-chain, a similar approach has already been proposed. In \cite{soares2013integrated} it is possible to see the tool-chain JADE+TraSMAPI+SUMO. However, the goal of this work was focused on the vehicles itself instead of traffic lights.

\section{Conclusions}
This paper explores the use of a specific tool-chain for the implementation of intelligent traffic light control. At the end, we have a tool that allows us to implement and test real MAS-based solutions in the domain of traffic and transportation, using commodity computers and open-source tools of wide reach. Q-Learning was used as the reinforcement learning method to illustrate the implementation of traffic light agents. The tool-chain resulting from the integration of JADE and SUMO through TraSMAPI is the main expected contribution of this paper.

Nonetheless, many improvements can be identified for future work. This paper did not analyse other forms for traffic control. For example, there are solutions based on the simple statistical analysis of traffic information and posterior adjustment according to such analytical procedures. This kind of solution can contrast with others as it can be highly dynamic and therefore can be applied to very specific scenarios. Another possible solution is the installation of sensors in each traffic light that, on a reactive way, can simply respond according to the number of waiting vehicles in the queue, needing neither great computation power nor the analysis of the traffic network, totally or partially.

The tool-chain itself could be improved in some different possible ways, including scalability, robustness, and efficiency. Firstly, the increase of performance in information retrieval by decreasing time in communication between the agent and the simulator. SUMO, that is still in its very young stage, proved to be much slower than desired with a larger number of vehicles and constant information retrieval. Certainly this aspect will be improved in next versions of SUMO, but it is necessary to analyse who is to blame: Is TraCI too much slow? is TraSMAPI implemented well in what concerns performance issues? During simulations is the number of generated requests to TraCI greater than necessary? and so forth! On other hand, for the real simulated system implementation it would be necessary to develop a distributed system where each agent was executed in each machine.

In this specific study we did not use JADE ability to distribute agents over a computer network, as our main objective is to demonstrate how JADE and SUMO can be integrated through TraSMAPI. Nonetheless, larger networks will certainly require more robust computational power, which can be achieved through an appropriate distribution of computation across a computer network. The traffic network itself could also be improved: a more realistic map for simulation can give more relevant results. Maybe the multi-agent system used could not be the best for the proposed approach. An analysis of the best tool to use is certainly imperative.

We intend to use the proposed framework to further investigate traffic control strategies through more robust and complex signal agents. Contrary to the manual approach adopted to set up semaphore plans, tools such as Transit can be used to assist a more coherent definition of phases at each junction of the network. Finally, in terms of general results, it seems that Q-Learning taking into account the duration of the phases and the period of the day obtains better general results, even if they are not very significant. Nevertheless, it is necessary to perform these experiments in more real settings, not only in what concerns the network, but also in what concerns simulation. So, it would be possible to better conclude whether the Q-Learning implementation in traffic networks is an added value not only for drivers, but also for the system as a whole.
 
\section{Acknowledgments}
Authors greatly acknowledge invaluable contributions by Filipe Oliveira, who also worked directly on this project. We also thank Professor Eug\'{e}nio Oliveira and Dr. Henrique Lopes Cardoso for important suggestions and comments on the course of this work.

\bibliographystyle{abbrv}
\bibliography{sample}

\begin{thebibliography}{10}

\bibitem{arel2010reinforcement}
I.~Arel, C.~Liu, T.~Urbanik, and A.~Kohls.
\newblock Reinforcement learning-based multi-agent system for network traffic
  signal control.
\newblock {\em Intelligent Transport Systems, IET}, 4(2):128--135, 2010.

\bibitem{azimirad2010novel}
E.~Azimirad, N.~Pariz, and M.-B.~N. Sistani.
\newblock {A} novel fuzzy model and control of single intersection at urban
  traffic network.
\newblock {\em Systems Journal, IEEE}, 4(1):107--111, 2010.

\bibitem{balmer2008agent}
M.~Balmer, K.~Meister, M.~Rieser, K.~Nagel, K.~W. Axhausen, K.~W. Axhausen, and
  K.~W. Axhausen.
\newblock {\em Agent-based simulation of travel demand: Structure and
  computational performance of MATSim-T}.
\newblock ETH, Eidgen{\"o}ssische Technische Hochschule Z{\"u}rich, IVT
  Institut f{\"u}r Verkehrsplanung und Transportsysteme, 2008.

\bibitem{balmer2009matsim}
M.~Balmer, M.~Rieser, K.~Meister, D.~Charypar, N.~Lefebvre, K.~Nagel, and
  K.~Axhausen.
\newblock {MATSim-T}: Architecture and simulation times.
\newblock {\em Multi-agent systems for traffic and transportation engineering},
  pages 57--78, 2009.

\bibitem{bazzan2010learning}
A.~L. Bazzan, D.~de~Oliveira, and B.~C. da~Silva.
\newblock Learning in groups of traffic signals.
\newblock {\em Engineering Applications of Artificial Intelligence},
  23(4):560--568, 2010.

\bibitem{bazzan2010itsumo}
A.~L. Bazzan, M.~d.~B. do~Amarante, T.~Sommer, and A.~J. Benavides.
\newblock {ITSUMO}: an agent-based simulator for its applications.
\newblock In {\em Proc. of the 4th Workshop on Artificial Transportation
  Systems and Simulation. IEEE}, 2010.

\bibitem{bazzan2013review}
A.~L. Bazzan and F.~Kl{\"u}gl.
\newblock A review on agent-based technology for traffic and transportation.
\newblock {\em The Knowledge Engineering Review}, pages 1--29, 2013.

\bibitem{chin2012q}
Y.~K. Chin, W.~Y. Kow, W.~L. Khong, M.~K. Tan, and K.~T.~K. Teo.
\newblock {Q}-learning {T}raffic {S}ignal {O}ptimization within {M}ultiple
  {I}ntersections {T}raffic {N}etwork.
\newblock In {\em Computer Modeling and Simulation (EMS), 2012 Sixth UKSim/AMSS
  European Symposium on}, pages 343--348. IEEE, 11 2012.

\bibitem{da2006itsumo}
B.~C. da~Silva, A.~L. Bazzan, G.~K. Andriotti, F.~Lopes, and D.~de~Oliveira.
\newblock {ITSUMO}: an intelligent transportation system for urban mobility.
\newblock In {\em Innovative Internet Community Systems}, pages 224--235.
  Springer, 2006.

\bibitem{de2006reinforcement}
D.~de~Oliveira, A.~L. Bazzan, B.~C. da~Silva, E.~W. Basso, L.~Nunes,
  R.~Rossetti, E.~de~Oliveira, R.~da~Silva, and L.~Lamb.
\newblock {R}einforcement {L}earning based {C}ontrol of {T}raffic {L}ights in
  {N}on-stationary {E}nvironments: {A} {C}ase {S}tudy in a {M}icroscopic
  {S}imulator.
\newblock In {\em EUMAS}. Citeseer, 2006.

\bibitem{de2010multi}
L.~B. de~Oliveira and E.~Camponogara.
\newblock Multi-agent model predictive control of signaling split in urban
  traffic networks.
\newblock {\em Transportation Research Part C: Emerging Technologies},
  18(1):120--139, 2010.

\bibitem{sumo:glance}
{German Aerospace Center, Institute of transportation Systems}.
\newblock {SUMO} at a {G}lance.
\newblock \url{http://sumo-sim.org/userdoc/Sumo_at_a_Glance.html}.
\newblock Accessed: 2013-10-23.

\bibitem{koonce2008traffic}
P.~Koonce, L.~Rodegerdts, K.~Lee, S.~Quayle, S.~Beaird, C.~Braud, J.~Bonneson,
  P.~Tarnoff, and T.~Urbanik.
\newblock Traffic signal timing manual.
\newblock Technical report, US Department of Transportation, 2008.

\bibitem{lu2011multi}
W.~Lu, Y.~Zhang, and Y.~Xie.
\newblock A multi-agent adaptive traffic signal control system using swarm
  intelligence and neuro-fuzzy reinforcement learning.
\newblock In {\em Integrated and Sustainable Transportation System (FISTS),
  2011 IEEE Forum on}, pages 233--238. IEEE, 2011.

\bibitem{moya2007towards}
L.~J. Moya and A.~Tolk.
\newblock Towards a taxonomy of agents and multi-agent systems.
\newblock In {\em Proceedings of the 2007 spring simulation
  multiconference-Volume 2}, pages 11--18. Society for Computer Simulation
  International, 2007.

\bibitem{russell2010artificial}
S.~Russell and P.~Norvig.
\newblock {\em {A}rtificial {I}ntelligence: {A} {M}odern {A}pproach}.
\newblock Prentice Hall series in artificial intelligence. Prentice Hall, 2010.

\bibitem{soares2013integrated}
G.~Soares, J.~Macedo, Z.~Kokkinogenis, and R.~J. Rossetti.
\newblock An integrated framework for multi-agent traffic simulation using sumo
  and jade.
\newblock In {\em SUMO2013, The first SUMO User Conference, May 15-17, 2013 -
  Berlin-Adlershof, Germany}, pages 125--131. DLR - Institut f{\"u}r
  Verkehrssystemtechnik, 2013.

\bibitem{teklu2007genetic}
F.~Teklu, A.~Sumalee, and D.~Watling.
\newblock {A} genetic algorithm approach for optimizing traffic control signals
  considering routing.
\newblock {\em Computer-Aided Civil and Infrastructure Engineering},
  22(1):31--43, 2007.

\bibitem{jade:website}
{Telecom Italia Lab}.
\newblock {JADE} description.
\newblock \url{http://jade.tilab.com/description-index.htm}.
\newblock Accessed: 2013-10-20.

\bibitem{Timoteo2012157}
I.~J. Tim\'{o}teo, M.~R. Ara\'{u}jo, R.~J. Rossetti, and E.~C. Oliveira.
\newblock Using trasmapi for the assessment of multi-agent traffic management
  solutions.
\newblock {\em Progress in Artificial Intelligence}, 1(2):157--164, 2012.

\bibitem{trasmapi:multisimul}
I.~J. P.~M. Tim\'{o}teo, M.~R. Ara\'{u}jo, R.~J.~F. Rossetti, and E.~C.
  Oliveira.
\newblock {T}ra{SMAPI}: {A}n {API} oriented towards {M}ulti-{A}gent {S}ystems
  real-time interaction with multiple {T}raffic {S}imulators.
\newblock In {\em Intelligent Transportation Systems (ITSC), 2010 13th
  International IEEE Conference on}, pages 1183--1188, 9 2010.

\bibitem{Vilarinho2014262}
C.~Vilarinho, G.~Soares, J.~Macedo, J.~P. Tavares, and R.~J. Rossetti.
\newblock Capability-enhanced \{AIMSUN\} with real-time signal timing control.
\newblock {\em Procedia - Social and Behavioral Sciences}, 111(0):262 -- 271,
  2014.
\newblock Transportation: Can we do more with less resources? - 16th Meeting of
  the Euro Working Group on Transportation - Porto 2013.

\bibitem{wegener2008traci}
A.~Wegener, M.~Pi{\'o}rkowski, M.~Raya, H.~Hellbr{\"u}ck, S.~Fischer, and J.-P.
  Hubaux.
\newblock {T}ra{CI}: an interface for coupling road traffic and network
  simulators.
\newblock In {\em Proceedings of the 11th communications and networking
  simulation symposium}, pages 155--163. ACM, 2008.

\end{thebibliography}

\end{document}